\documentclass[aps,prl,twocolumn]{revtex4}
\usepackage{amsfonts}
\usepackage{amsmath}
\usepackage{amssymb}
\usepackage{graphicx}

\begin{document}
\preprint{ }
\title{Antisymmetric magnetoresistance of the ${\rm SrTiO_3}/{\rm LaAlO_3}$ interface}
\author{Snir Seri}
\author{Lior Klein}
\affiliation{Department of Physics, Nano-magnetism Research Center, Institute of Nanotechnology and Advanced Materials, Bar-Ilan University, Ramat-Gan 52900, Israel}


\begin{abstract}

The longitudinal resistance $R_{xx}$ of the ${\rm SrTiO_3}/{\rm LaAlO_3}$ interface with magnetic fields applied perpendicular to the interface has an antisymmetric term (namely, $R_{xx}(H)\neq R_{xx}(-H)$) which  increases with decreasing temperature and increasing field. We argue that the origin of this phenomenon is a non-homogeneous  Hall effect with clear contribution of an extraordinary Hall effect, suggesting the presence of non-uniform field-induced magnetization.

\end{abstract}




\maketitle

The quasi-two-dimensional electron gas (q2DEG) that forms at the interface between the two insulating oxides, ${\rm SrTiO_3}$ and ${\rm LaAlO_3}$ (LAO/STO), has fascinated  many researchers who have been trying to elucidate the properties of this system \cite{high mobility,tunable quasi,superconductivity2,origin unusual final,origin perspectives,Mapping,complementary,oxygen vacancies,LAO STO}. Nevertheless, some of the most basic properties of this material are still controversial. Thus, contrary to the apparent consensus concerning the superconducting ground state of this system below 300 mK which obeys the Kosterlitz-Thouless phase transition \cite{superconductivity2,superconductivity1}, the existence and nature of the magnetism is still an open question. While there is a  theoretical prediction for a magnetic order \cite{magnetic Theory2}, the experimental situation is more complicated where some groups report hysteretic magnetoresistance which suggests ferromagnetic order \cite{ferromagnetism} and other report lack of hysteresis but unusual magnetoresistance behavior which they attribute to some kind of magnetic order \cite{antiferromagnetism}.

Here, we present data showing that the magnetoresistance (MR) of the  LAO/STO interface with magnetic fields applied perpendicular to the interface has an antisymmetric term which increases with decreasing temperature and increasing field. While the qualitative behavior is common to all the patterns we have studied, the magnitude and the sign of the phenomenon vary considerably even between neighboring segments of the same pattern. Based on field, temperature and angular dependent measurements of the Hall effect (HE) and the MR, we argue that the likely source of this phenomenon is a non-homogeneous HE with a clear contribution of a non-uniform extraordinary Hall effect (EHE) \cite{AHE}. This interpretation implies that the applied magnetic field induces non-uniform magnetization. The non-uniform field-induced magnetization may suggest that either the induced magnetization is extrinsic to the q2DEG or that other non-uniformity affects locally the electron gas magnetization. The induced magnetization is likely to be the source of the observed large positive and negative MR when magnetic fields are applied perpendicular and parallel to the interface, respectively. The negative MR is in the form of sharp and narrow dips, indicating strong magnetic anisotropy.

\begin{figure}[ht]
\includegraphics[scale=0.35, trim=100 0 100 0]{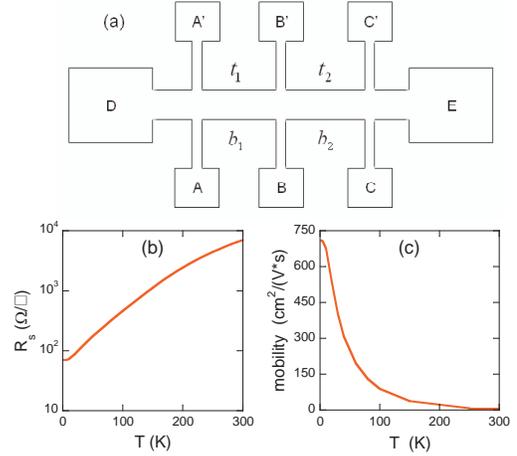}
\caption{(a) A sketch of the pattern. The width of the current path (between D and E) is 100 $\mu m$. The distance between neighboring Hall crosses is 300 $\mu m$. (b) The sheet resistance as a function of temperature. (c) The mobility as a function of temperature.} \label{property}
\end{figure}

While we do not resolve the elusive issue of magnetism in LAO/STO interfaces, we present multiple pieces of evidence for non-uniform field-induced magnetization at low temperatures which provide central ingredients for elucidating the nature of the q2DEG. The evidence for sizable EHE which we use for detecting the magnetism is of importance by itself for exploring the transport mechanism in this system and in addition it opens the door for future spintronics applications as it may be used for detecting spin injection into the q2DEG \cite{tunable quasi,magnetoelectronics}.

The growth and patterning method of our samples, which were provided by the Augsburg group, have been reported elsewhere \cite{microlithography}. For this study we use patterns as shown in Figure \ref{property} with current paths that are 50 microns or 100 microns wide and three pairs of voltage leads that allow for simultaneous longitudinal and transverse voltage measurements. The figure also shows the temperature dependence of the sheet resistance and the mobility of the sample whose data are presented here. The data are similar to those reported previously for samples with conductivity dominated by intrinsic interface doping \cite{high mobility,origin unusual final,complementary,oxygen vacancies,ferromagnetism}.


We measured the sheet resistance for each pattern in four different locations: two on the bottom side of the current path ($R_{b1}$ and $R_{b2}$) and  two on the upper side of the current path ($R_{t1}$ and $R_{t2}$) (see Fig. 1a). Surprisingly, as we measured the MR of the four segments at low temperatures, we noticed that when the field polarity was reversed, the resistivity slightly changed (usually by few percents) which implies that the longitudinal resistance $R_{xx}$ has symmetric and antisymmetric components: $R_{xx}^S$ and  $R_{xx}^{AS}$ defined as  $R_{xx}^S= (R_{xx}(H)+R_{xx}(-H))/2$ and $R_{xx}^{AS}= (R_{xx}(H)-R_{xx}(-H))/2$. The existence of an antisymmetric component was also noticed when we exchanged the current and voltage leads without reversing the field  as expected by the reciprocity theorem \cite{reciprocity}. According to this theorem $R_{12,34}(H)=R_{34,12}(-H)$ where the first pair of indices represents the terminals used to supply and draw current, and the second pair of indices represents the terminals used to measure the potential difference. Therefore, $R^{S}_{12,34}(H)=(R_{12,34}(H)+R_{34,12}(H))/2$ and $R^{AS}_{12,34}(H)=(R_{12,34}(H)-R_{34,12}(H))/2$.

\begin{figure}[ht]
\includegraphics[scale=0.47, trim=100 0 100 0]{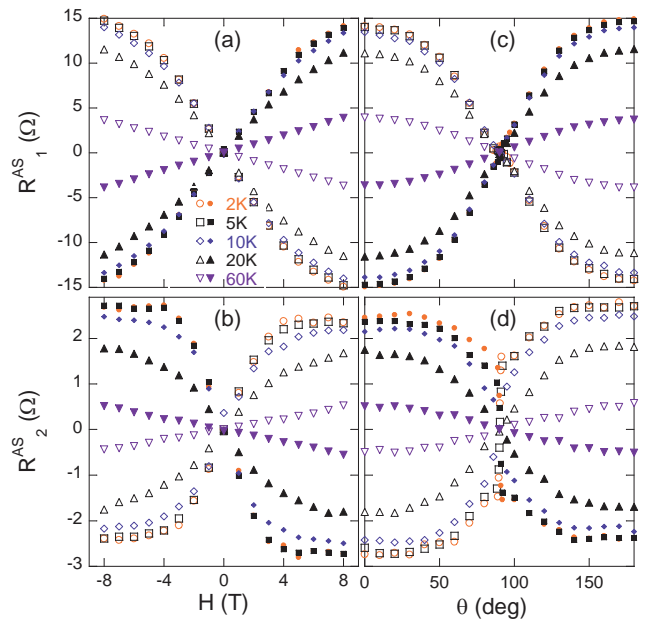}
\caption{$R^{AS}$ of the pair $b1-t1$ ((a) and (c)) and $b2-t2$ ((b) and (d)) as a function of  a magnetic field applied perpendicular to the interface at different temperatures ((a) and (b)) and as a function of the angle $\theta$ between a field of ${\rm 8 \ T}$ and the normal to the interface at different temperatures ((c) and (d)). Open and full symbols are used for the bottom and upper sides of the current path, respectively.}
\label{Ras}
\end{figure}

Figure \ref{Ras} shows the antisymmetric components ($R^{AS}$) of the longitudinal resistances as measured between different leads as a function of a magnetic field applied perpendicular to the interface, and for a constant magnetic field as a function of the angle $\theta$ between the field and the normal to the interface. Figure \ref{Ras}a and Figure \ref{Ras}c show the antisymmetric component of the resistances measured between the bottom leads A and B ($R^{AS}_{b1}$) and the top leads A' and B' ($R^{AS}_{t1}$) which we denote the $b1-t1$ pair. Figure \ref{Ras}b and Figure \ref{Ras}d show the antisymmetric component of the resistances measured between the bottom leads B and C ($R^{AS}_{b2}$) and the top leads B' and C' ($R^{AS}_{t2}$) which we denote the $b2-t2$ pair. We see that for each pair,  the antisymmetric component is reversed; namely, $R^{AS}_{b1}(H)=-R^{AS}_{t1}(H)$ or $R^{AS}_{b1}(H)=R^{AS}_{t1}(-H)$. On the other hand, there are very significant differences between the two pairs despite the fact that they are both part of the same pattern. The sign of the signal is reversed, its magnitude is different ($\sim15 \ \Omega$  in the $b1-t1$ pair compared to less than $3 \ \Omega$  in the $b2-t2$ pair at ${\rm 2 \ K}$ with a field of ${\rm 8 \ T}$) and the angular dependence is quantitatively different where  $R^{AS}$ of the $b2-t2$ pair exhibits at low temperatures sharp jumps around $\theta=90^{\circ}$. In addition, the change with temperature is much bigger for the $b2-t2$ pair. At the same time, we note that the variations in the symmetric component of the sheet resistance as measured between different leads of a pattern do not exceed 1 percent.

Antisymmetric contributions to longitudinal MR were observed before in other systems of two dimensional electron gas and were attributed to variations in the HE that may arise in these systems due to carrier density gradients \cite{gradients}. When such a gradient exists, it induces a directional change in the HE and thus it contributes to a change in the voltage drop measured between two points on the same side of the current path. Therefore, the antisymmetric contribution behaves as the difference between the HE signals with different carrier densities.

In our case, when a magnetic field induces $V_A(H)-V_{A'}(H)=\Delta_1$ and $V_B(H)-V_{B'}(H)=\Delta_2$, a term equal to $(\Delta_1-\Delta_2)/2$  adds to $R_{b1}$ and a term equal to $(\Delta_2-\Delta_1)/2$ adds to $R_{t1}$. These terms which add to the longitudinal resistances are the antisymmetric components and we thus expect that $R^{AS}_{b1}=-R^{AS}_{t1}$, as indeed measured (some deviations from this relation can occur due to misalignment of the voltage leads) \cite{note1}.

While we expect the antisymmetric contribution to behave qualitatively like the HE, in fact, in some cases the behavior is strikingly different. We refer particularly to the behavior of $R^{AS}_{b2}$ and $R^{AS}_{t2}$ which show at low temperatures saturation with field (Fig. \ref{Ras}b) and abrupt sign reversal when the magnetic field changes the polarity of its perpendicular component (Fig. \ref{Ras}d).  This kind of behavior cannot be reconciled with a contribution of a purely  ordinary Hall effect (OHE) and thus we are drawn to the conclusion that there is a contribution of an EHE  which is sensitive to the perpendicular component of the local magnetic moments.

\begin{figure}[ht]
\includegraphics[scale=0.48, trim=100 0 100 0]{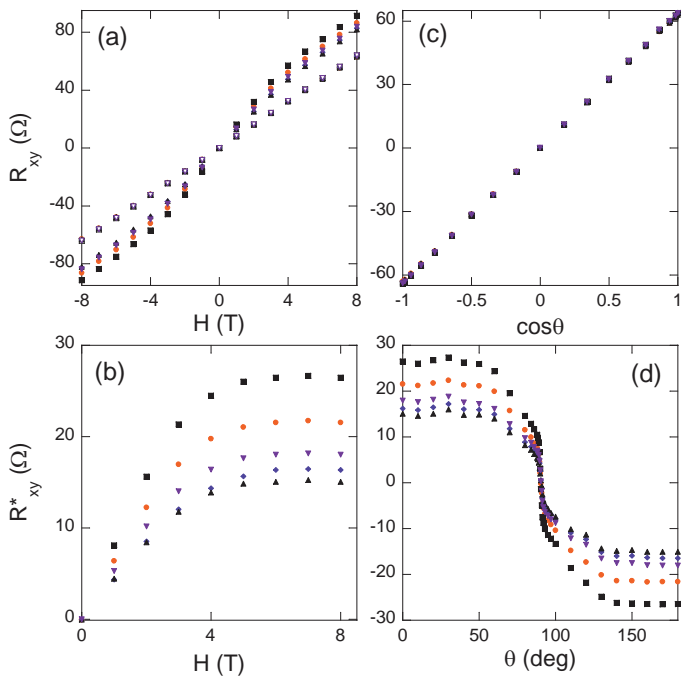}
\caption{(a) The HE measurements as a function of a magnetic field on five different crosses of the same sample. Full and open symbols are used for temperature of ${\rm 2 \ K}$ and ${\rm 60 \ K}$, respectively. (b) The HE after subtracting the assumed OHE extracted from the high field limit ($R^{*}_{xy}$) for the same five crosses at ${\rm 2 \ K}$. (c) The HE measurements as a function of cos$\theta$, where $\theta$ is the angle between a field of ${\rm 8 \ T}$ and the normal to the interface for the same five crosses at ${\rm 60 \ K}$. (d) $R^{*}_{xy}$ as a function of the angle $\theta$ for the same five crosses at ${\rm 2 \ K}$ and a field of ${\rm 8 \ T}$.} \label{HE4}
\end{figure}

Figure \ref{HE4}a shows HE measurements performed at ${\rm 2 \ K}$  on five different crosses of the same sample.  All the crosses show a similar linear behavior at high fields; however, at low fields there is a different degree of non-linearity. As temperature is increased, the spread of the HE curves practically disappears, the HE is linear in field (see Fig. \ref{HE4}a) and its angular dependence is as expected from the OHE (see Fig. \ref{HE4}c). It thus appears that the HE has two contributions: an OHE which is linear in field and is weakly temperature dependent and an EHE which saturates with field at low temperature and its magnitude varies strongly with temperature. Figure \ref{HE4}b shows the HE after subtracting the assumed OHE extracted from the high field limit (denoted $R^{*}_{xy}$). We see for the five crosses a qualitatively similar saturating behavior although the magnitude varies. The angular dependence of the non-linear term exhibits the sharp features exhibited by $R^{AS}$ and here again we see the abrupt changes when the perpendicular component of the field changes its polarity.

\begin{figure}[ht]
\includegraphics[scale=0.46, trim=100 0 100 0]{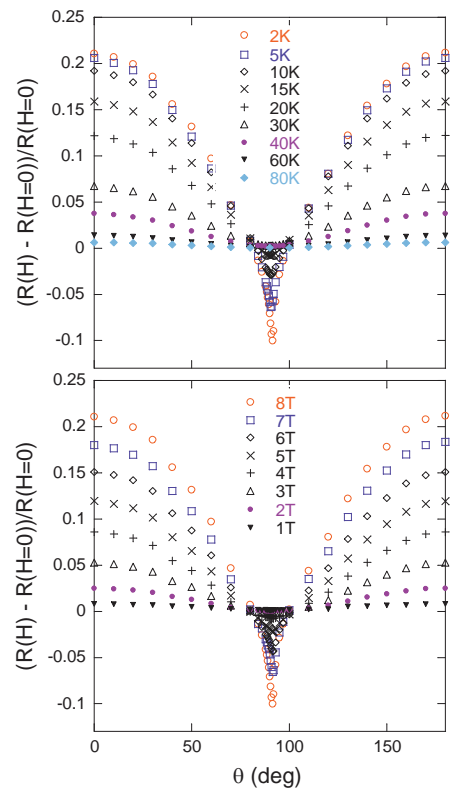}
\caption{The magnetoresistance as a function of $\theta$ with a field of ${\rm 8 \ T}$ at different temperatures (Top) and at ${\rm T=2 \ K}$ with different fields (Bottom).} \label{MR}
\end{figure}

We see (Fig. \ref{HE4}a) that the spread in the high field slopes (which we relate to the OHE) is much smaller than the spread in the non-linear contribution which saturates at high fields and we attribute to the EHE.
When we measure the antisymmetric component of the longitudinal resistivity we obtain  the difference in the HE at two locations and due to the smaller spread in the OHE  we see in $R^{AS}$ a more dominant contribution of the EHE. This is particularly true for $R^{AS}_{b2}$ and $R^{AS}_{t2}$.

Figs. \ref{Ras} and \ref{HE4} show that $R^{*}_{xy}$ and $R^{AS}$ (when it is dominated by changes in EHE) are almost constant up to $\theta\sim45^{\circ}$ and they change abruptly around $\theta=90^{\circ}$. This behavior suggests that the induced magnetization does not follow the orientation of the applied field - it remains closer to the perpendicular direction until it abruptly switches its perpendicular polarity near $\theta=90^{\circ}$. This behavior indicates the existence of magnetic anisotropy with an easy axis perpendicular to the film plane.

The abrupt change in magnetic orientation is clearly manifested in the MR. Figure \ref{MR} shows the MR as a function of $\theta$ either with a field of ${\rm 8 \ T}$ at different temperatures or at ${\rm 2 \ K}$ with different fields. We see that with decreasing temperature and/or increasing magnetic field there is a large increase in the positive MR at $\theta=0^{\circ}$ and a large increase in the negative MR at $\theta=90^{\circ}$; however, the two peaks are very different and the negative peak at $\theta=90^{\circ}$ is very narrow \cite{antiferromagnetism}. As the MR depends on the angle between the magnetization and the film plane, we can attribute the narrowness of the negative peaks to the abrupt change in the magnetization orientation observed in the HE measurements. We also note that the angular dependence of the MR has the same behavior weather  the plane perpendicular to the film plane in which the field is rotating is perpendicular to the current path or it includes the current path. Namely, the effect is not a regular anisotropic MR effect. At this point it is unclear why the change in magnetic orientation has such a dramatic effect on the transport properties of the q2DEG.

While we do not completely exclude a non-magnetic scenario for the HE behavior, our observations suggest that the HE is composed of a relatively uniform contribution of an OHE and a very non uniform EHE which originates from field-induced magnetization where the magnetization decreases with temperature and exhibits perpendicular magnetic anisotropy. The magnetic scenario is consistent with the saturation behavior typical of magnetism exhibited by the difference between HE measurements performed at different points on the same sample (Fig. \ref{Ras}). It is consistent with the observation that the HE measured at various points is very similar at high temperatures (60 K) whereas at low temperatures it maintains the same linear slope at high fields while the non-linear contribution varies considerably from point to point (Fig. \ref{HE4}). In addition, it is consistent with the angular dependence of the non-linear contribution of the HE and the correlation with the angular dependence of the MR (Fig. \ref{MR}).

What is the nature of the magnetism and what is its origin? First, we note that in all our measurements we have not observed  hysteretic effects. Therefore, there is no direct evidence for spontaneous magnetization.  We also see that the induced magnetization is non-uniform which raises the question whether the magnetization is an intrinsic property of the q2DEG itself or the magnetization is extrinsic and it only affects the q2DEG. To explain the non-uniform magnetization in the intrinsic scenario we would need to assume that external changes affect the q2DEG and vary its magnetic response locally. The other scenario would be that there are magnetic impurities  at or near the interface that give rise to the magnetic effects. The fact that we see no clear correlation between changes in carrier density and changes in magnetic moment density (as deduced from the qualitative variations  of $R^{AS}$) may suggest an extrinsic scenario. On the other hand, we do observe strong magnetic anisotropy; namely, it is easier to magnetize perpendicular to the interface, which could be a signature of the two dimensional nature of the magnetism; hence, it may point towards the intrinsic scenario.

L. K. acknowledges support by the Israel Science Foundation founded by the Israel Academy of Sciences and Humanities (Grant no. 577/07) and by the German Israeli Foundation (Grant no. 979/2007). We acknowledge the contribution of J. Mannhart and S. Thiel who provided the samples used for this research. We acknowledge useful discussions with K. A. Moler, E. Shimshoni and Y. Strelniker.


\begin{thebibliography}{9}
\bibitem{note1} We also note that the symmetric components of the resistances measured on the two sides of the current path may have a slightly different field dependence related to local variations in the resistivity and therefore there could be differences between the antisymmetric contribution and $R_{b1}-R_{t1}$.
\end{thebibliography}
\end{document}